\documentclass[12pt]{article}
\usepackage{epsfig}
\usepackage{color}
\usepackage{amssymb,amsmath}

\newcommand{\bea}{\begin{eqnarray}}
\newcommand{\eea}{\end{eqnarray}}
\newcommand{\be}{\begin{equation}}
\newcommand{\ee}{\end{equation}}

 \makeatletter
\def\alt{\mathrel{\mathpalette\gl@align<}}
\def\agt{\mathrel{\mathpalette\gl@align>}}
\def\gl@align#1#2{\lower.6ex\vbox{\baselineskip\z@skip\lineskip\z@
\ialign{$\m@th#1\hfil##\hfil$\crcr#2\crcr\sim\crcr}}} \makeatother

\begin{document}
\begin{flushright}
KEK-TH-1422
\end{flushright}
\vspace*{1.0cm}

\begin{center}
\baselineskip 20pt 
{\Large\bf 
Resonant Leptogenesis in the Minimal $B-L$ Extended Standard Model at TeV  
}
\vspace{1cm}

{\large 
Satoshi Iso$^{a,}$\footnote{satoshi.iso@kek.jp}, 
Nobuchika Okada$^{b,}$\footnote{okadan@ua.edu} 
and 
Yuta Orikasa$^{a,}$\footnote{orikasa@post.kek.jp}
} \vspace{.5cm}

{\baselineskip 20pt \it
$^{a}$ KEK Theory Center, \\ 
High Energy Accelerator Research Organization (KEK)  \\
and \\
Department of Particles and Nuclear Physics, \\
The Graduate University for Advanced Studies (SOKENDAI), 
\\
1-1 Oho, Tsukuba, Ibaraki 305-0801, Japan  \\
\vspace{3mm}
$^{b}$ Department of Physics and Astronomy, 
University of Alabama, 
Tuscaloosa,  AL 35487, USA}

\vspace{.5cm}

\vspace{1.5cm} {\bf Abstract}\\
\end{center}

We investigate the resonant leptogenesis scenario 
 in the minimal $B-L$ extended standard model (SM) 
 with the $B-L$ symmetry breaking at the TeV scale. 
Through detailed analysis of the Boltzmann equations, 
 we show how much the resultant baryon asymmetry 
 via leptogenesis is enhanced or suppressed, 
 depending on the model parameters, in particular, 
the neutrino Dirac Yukawa couplings and the
 TeV-scale Majorana masses of heavy degenerate neutrinos. 
In order to consider a realistic case, 
 we impose a simple ansatz for the model parameters 
 and analyze the neutrino oscillation parameters 
 and the baryon asymmetry via leptogenesis 
 as a function of only a single CP-phase. 
We find that for a fixed CP-phase 
 all neutrino oscillation data and the observed baryon 
 asymmetry of the present universe 
 can be simultaneously reproduced. 

\thispagestyle{empty}

\newpage

\addtocounter{page}{-1}
\setcounter{footnote}{0}
\baselineskip 18pt
\section{Introduction}

The origin of the baryon asymmetry in the present universe 
 is one of the big mysteries in cosmology. 
The ratio of the baryon (minus anti-baryon) density 
 $n_B$ to the entropy density $s$  has been measured 
 with the precision at 10\% level by the WMAP satellite 
 experiment \cite{WMAP}, 
\bea 
 Y_B=\frac{n_B}{s}=0.87\times10^{-10}.
\eea 
It might be the most attractive if the origin of the baryon asymmetry  
 can be explained within the context of the SM, 
 the electroweak baryogenesis \cite{EWB}. 
In order for this scenario to work, a strong first  order 
 electroweak phase transition is necessary in the early universe. 
However, the Higgs potential satisfying the current lower bound 
 on the SM Higgs boson mass \cite{HiggsMass} 
 is not likely to show this strong first 
 order phase transition and hence, the SM electroweak baryogenesis 
 is almost ruled out. 

An appealing alternative is the leptogenesis scenario \cite{leptogenesis}, 
 which is also intimately related with the smallness 
 of the neutrino masses through the seesaw mechanism \cite{seesaw}. 
A most widely accepted scenario is to extend the SM 
 by introducing the right-handed Majorana neutrinos with masses
 around an intermediate scale whose out-of-equilibrium decays 
 create lepton asymmetry in the universe. 
The lepton asymmetry is converted to the baryon asymmetry
 through the $(B+L)$-violating sphaleron transitions 
 \cite{sphaleron1, sphaleron2} 
 with the conversion rate \cite{Khlebnikov}
\bea 
 Y_B =-\frac{8N_f+4N_H}{22N_f+13N_H}Y_L 
     =-\frac{28}{79}Y_L , 
\label{eq:blcomv}
\eea 
where we have taken $N_f=3$ and $N_H=1$ are the numbers of fermion families and 
 Higgs doublets in the SM. 
In normal thermal leptogenesis, 
 there is a lower bound on the mass of Majorana neutrinos 
 $\gtrsim 10^{10}$ GeV \cite{BBP} in order to create 
 sufficient amount of the baryon asymmetry. 
If it is the case, it is hopeless to directly observe 
 the heavy neutrinos at high-energy colliders 
 in the near future.

Many models beyond the SM have been proposed, 
 which may be realized at the TeV scale and hence accessible 
 to the Large Hadron Collider (LHC) currently 
 in operation and more future colliders 
 such as the International Linear Collider. 
Among many models, in this paper, we consider the minimal gauged 
 $B-L$ extended SM. 
This is an elegant and simple extension of the SM, 
 in which the right-handed neutrinos of three generations 
 are necessarily introduced for the cancellation
 of the gauge and gravitational anomalies. 
In addition, the mass of right-handed neutrinos arises 
 associated with the U(1)$_{B-L}$ gauge symmetry breaking 
 and the seesaw mechanism  is automatically implemented. 
In the view point of LHC physics, it is very interesting 
 if the $B-L$ symmetry breaking scale lies around TeV 
 so that the $B-L$ gauge boson (Z' boson) and the right-handed 
 neutrinos can be discovered in the near future \cite{BL-LHC}. 
Recently, we have proposed the minimal $B-L$ model 
 with the classical conformal invariance \cite{IOO1} 
 and showed that the $B-L$ symmetry breaking in this model 
 is naturally realized at the TeV scale 
 when the $B-L$ gauge coupling constant is the same order of 
 magnitude as the size of the SM gauge coupling constants \cite{IOO2}.

Although the minimal $B-L$ model at TeV is a very attractive 
 scenario, the normal thermal leptogenesis scenario cannot work 
 because the mass scale of the right-handed neutrinos is far below 
 the bound, $10^{10}$ GeV mentioned above. 
In this case, the CP-asymmetry parameter, which is roughly 
 proportional to Dirac Yukawa coupling squareds 
 is too small to give sufficient amount of baryon asymmetry 
 in the universe. 
However, it has been found that when two right-handed neutrinos 
 have almost degenerate masses, there is an enhancement 
 of the CP-asymmetry parameter \cite{RLG1}, 
 and this enhancement can make the leptogenesis scenario 
 viable even if the mass scale of the right-handed neutrinos 
 lie around TeV, the resonant leptogenesis \cite{RLG}. 
The maximum enhancement is achieved when the mass splitting 
 between two right-handed neutrinos is comparable to 
 the decay width of either right-handed neutrinos. 
By tuning the mass splitting between two right-handed neutrinos, 
 even a CP-asymmetry parameter of order unity can be obtained 
 in principle. 
However, it is still non-trivial whether the minimal $B-L$ model 
 at the TeV scale can reproduce the observed baryon asymmetry 
 because, as we will discuss later in detail, 
 the creation of the lepton asymmetry via decays 
 of right-handed neutrinos is highly suppressed 
 in the presence of the U(1)$_{B-L}$ gauge interaction 
 with Z' boson mass at the TeV scale \cite{Rabi}.

In this paper, we investigate in detail 
 the resonant leptogenesis scenario 
 in the minimal $B-L$ extended SM 
 with the $B-L$ symmetry breaking at the TeV scale. 
Through detailed analysis of the Boltzmann equations 
 with a variety of model-parameter sets, 
 we show how much the resultant baryon asymmetry 
 via leptogenesis is enhanced or suppressed, 
 depending on model parameters, in particular, 
 neutrino Dirac Yukawa couplings and 
 TeV-scale Majorana masses of heavy degenerate neutrinos. 
In order to consider a realistic case, 
 we impose a simple ansatz for model parameters 
 and analyze the neutrino oscillation parameters 
 and the baryon asymmetry via leptogenesis 
 as a function of only a single CP-phase. 
We find that a fixed CP-phase can simultaneously reproduce 
 all neutrino oscillation data and the observed baryon 
 asymmetry in the present universe.

The paper is organized as follows. 
In the next section, we give a brief review on 
 the minimal $B-L$ model and the natural realization 
 of the $B-L$ symmetry breaking at the TeV scale. 
In section 3, we analyze in detail the resonant leptogenesis 
 at the TeV scale by numerically solving the Boltzmann equations 
 with various parameter sets. 
We show how the generated baryon asymmetry depends on 
 the model parameters such as Dirac Yukawa coupling, 
 right-handed neutrino mass spectrum, etc. 
In section 4, we investigate more realistic parameter choices 
 so as to reproduce the neutrino oscillation data. 
We introduce two right-handed neutrinos 
 and a simple ansatz among the parameters,  
 by which the neutrino oscillation parameters and 
 the baryon asymmetry are determined by only 
 a single CP-phase. 
We find that there exists a CP-phase which 
 simultaneously reproduces the neutrino oscillation data 
 and the observed baryon asymmetry. 
The last section is devoted for conclusions. 
Formulas used in our analysis are listed in Appendix.

\section{The Minimal $B-L$ Model at TeV}

The minimal $B-L$ extended SM is based on 
 the gauge group 
 SU(3)$_c\times$SU(2)$_L \times $U(1)$_Y \times$U(1)$_{B-L}$ 
 with the particle contents listed in Table 1. 
The right-handed neutrinos ($N_i$) of three generations 
 are necessarily introduced by which all the gauge and 
 gravitational anomalies are canceled. 
The SM singlet scalar field ($\Phi$) works to break 
 the U(1)$_{B-L}$ gauge symmetry by 
 its vacuum expectation value (VEV), 
 $\langle \Phi \rangle =v_{B-L}/\sqrt{2}$. 
Once the $B-L$ gauge symmetry is broken, 
 the Z' boson acquires mass, 
\be
  m_{Z'} = 2 g_{B-L} v_{B-L},  
\ee
 where $g_{B-L}$ is the $B-L$ gauge coupling. 
The current experimental bound was found to be 
 $v_{B-L} \gtrsim$ 3 TeV \cite{vBL}. 

\begin{table}[t]
\begin{center}
\begin{tabular}{c|ccc|c}
            & SU(3)$_c$ & SU(2)$_L$ & U(1)$_Y$ & U(1)$_{B-L}$  \\
\hline
$ q_L^i $    & {\bf 3}   & {\bf 2}& $+1/6$ & $+1/3$  \\ 
$ u_R^i $    & {\bf 3} & {\bf 1}& $+2/3$ & $+1/3$  \\ 
$ d_R^i $    & {\bf 3} & {\bf 1}& $-1/3$ & $+1/3$  \\ 
\hline
$ \ell^i_L$    & {\bf 1} & {\bf 2}& $-1/2$ & $-1$  \\ 
$ N_i$   & {\bf 1} & {\bf 1}& $ 0$   & $-1$  \\ 
$ e_R^i  $   & {\bf 1} & {\bf 1}& $-1$   & $-1$  \\ 
\hline 
$ H$         & {\bf 1} & {\bf 2}& $+1/2$  &  $ 0$  \\ 
$ \Phi$      & {\bf 1} & {\bf 1}& $  0$  &  $+2$  \\ 
\end{tabular}
\end{center}
\caption{
Particle content:  
In addition to the SM particles, 
 right-handed neutrinos $N_i$ 
($i=1,2,3$ denotes the generation index) 
 and a complex scalar $\Phi$ are introduced. 
}
\end{table}

The Lagrangian relevant for the seesaw mechanism is given by
\bea 
  {\cal L} \supset -y_D^{ij} \overline{\nu_R^i} H \ell_L^j  
 - \frac{1}{2} y_N^i \Phi \overline{\nu_R^{i c}} \nu_R^i 
 +{\rm h.c.},  
\label{Yukawa}
\eea 
 where without loss of generality, we work on the basis 
 in which the second term is diagonalized and $y_N^i$ is 
 real and positive. 
The first term gives the Dirac neutrino mass term 
 after the electroweak symmetry breaking 
 ($m_D=y_D \langle H \rangle$), 
 while the right-handed neutrino Majorana masses 
 are generated through the second term associated 
 with the $B-L$ gauge symmetry breaking:  
\bea
 M_i = \frac{y_N^i}{\sqrt{2}} v_{B-L}. 
\eea

The $B-L$ symmetry breaking scale is determined by parameters 
 in the Higgs potential and in general it can be taken
 to be any scale as long as the experimental bound 
 $v_{B-L} \gtrsim 3$ TeV \cite{vBL} is satisfied. 
As discussed in the previous section, we assume 
 the $B-L$ symmetry breaking at the TeV scale in this paper,  
 and the masses of Z' boson and right-handed neutrinos 
 lie around TeV. 
In fact, it has been pointed out in \cite{IOO1, IOO2}
 if we impose the classical conformal symmetry 
 on the minimal $B-L$ model, the $B-L$ symmetry breaking 
 can be naturally realized at the TeV scale. 
In the rest of this section, we would like to 
 briefly review the classically conformal 
 $B-L$ extended standard model proposed in \cite{IOO1}. 
However, since the classical conformal invariance is not 
 important for the leptogenesis scenario 
 (except that it naturally leads to the TeV scale), 
 readers can skip to the next section.

We first note that because of its chiral nature, 
 the SM Lagrangian at the classical level possesses 
 the conformal invariance except for the Higgs mass term, 
 closely related to the gauge hierarchy problem. 
Bardeen has argued \cite{Bardeen} that 
 once the classical conformal invariance and its minimal violation 
 by quantum anomalies are imposed on the SM,
 it could be free from the quadratic divergences and 
 thus the gauge hierarchy problem. 
If the mechanism really works, we can directly interpolate 
 the electroweak scale and the Planck scale.
Since the classical conformal symmetry forbids the mass term 
 in the Higgs potential, the electroweak symmetry should be broken 
 radiatively through the Coleman-Weinberg (CW) mechanism \cite{CW}. 
Although this is an attractive scenario, 
 the effective Higgs potential is found to be unbounded 
 from below because of the large top Yukawa coupling 
 and therefore the classically conformal SM cannot be 
 a realistic scenario.

In \cite{IOO1}, we proposed a classically conformal minimal 
 $B-L$ model and showed that the $B-L$ gauge symmetry breaking 
 is successfully achieved via the CW mechanism and then, 
 this breaking triggers the electroweak symmetry breaking. 
Because of the CW mechanism, the SM singlet Higgs boson 
 associated with the $B-L$ symmetry breaking is much lighter 
 than Z' boson, 
\bea 
 \left( \frac{m_\phi}{m_{Z'}} \right)^2 
  \simeq \frac{6}{\pi} 
  \left( \alpha_{B-L} 
  - \frac{1}{96} \frac{\sum_i(\alpha_N^i)^2}{\alpha_{B-L}}
    \right)  \ll 1,
\label{phimass2}
\eea 
where $\alpha_{B-L} = g_{B-L}^2/(4 \pi)$, and 
      $\alpha_N^i =(y_N^i)^2/(4\pi)$. 
This formula also indicates the upper bound on $ \alpha_N^i $ 
 to keep the vacuum stability, $m_\phi^2 > 0$. 
Assuming a hierarchical Majorana mass spectrum, for example, 
 we find the upper bound on the heaviest right-handed neutrino mass as 
\bea 
\sum_im_{N_i}^4  <  \frac{3}{2} m_{Z'}^4.   
\label{RNmassbound}
\eea

Once the $B-L$ symmetry is broken, the Z' boson and 
 the right-handed neutrinos acquire masses 
 at the $B-L$ symmetry breaking scale. 
Their masses contribute to the effective mass of 
 the SM Higgs doublet through quantum corrections. 
The naturalness argument, namely, the quantum corrections 
 should not exceed the electroweak scale so far, 
 leads to an upper bound on the $B-L$ symmetry breaking scale. 
Two-loop corrections with Z' boson and top quarks are 
 found to be dominant, and we have concluded \cite{IOO2} 
 that the $B-L$ symmetry breaking scale should be around TeV.

\section{Leptogenesis in the minimal $B-L$ model at TeV}

Now we study baryogenesis via leptogenesis 
 in the minimal $B-L$ model at TeV. 
The lepton asymmetry in the universe is generated 
 by the CP-violating out-of-equilibrium decays 
 of right-handed neutrinos, and this asymmetry 
 is converted to the baryon asymmetry through 
 the sphaleron process with the conversion rate 
 $Y_B=-(79/28) Y_L$. 
The generated  baryon asymmetry is evaluated by  
 solving the Boltzmann equations. 
When the Majorana masses of three right-handed neutrinos 
 are largely different as usually assumed, 
 it is sufficient to consider the Boltzmann equations 
 only for the lightest right-handed neutrino. 
This is because the lepton asymmetry generated by 
 heavier right-handed neutrinos are washed out 
 by the inverse-decay of the lightest right-handed neutrinos 
 before its out-of-equilibrium decay \cite{boltz}. 
However, in the resonant leptogenesis scenario, 
 (at least) two right-handed neutrinos are degenerate in mass 
 and it is generally not clear whether analysis of 
 the Boltzmann equations with only one right-handed neutrino is sufficient. 
As we will see later, it can be essential for general cases 
 to consider the Boltzmann equations for multiple right-handed neutrinos.

\begin{figure}[t]
\leavevmode
\begin{center}
{\includegraphics[scale=1.0]{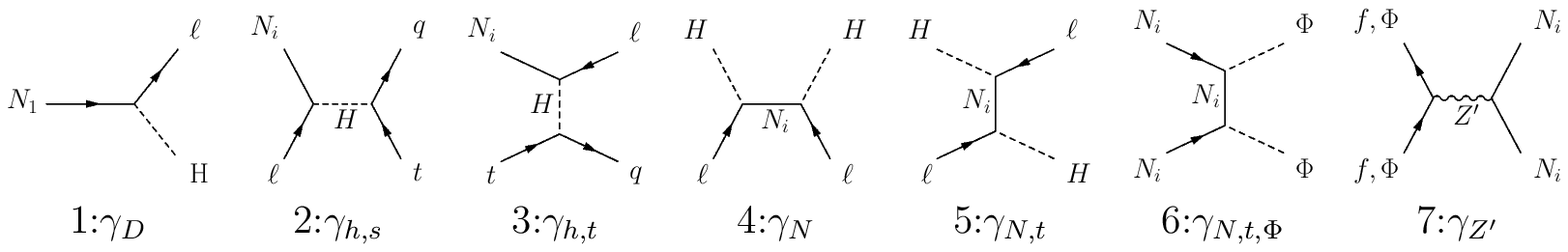}}
\caption{
Feynman diagrams corresponding to each $\gamma$s. 
}
\label{gamma}
\end{center}
\end{figure}

We begin our analysis with the Boltzmann equations 
 in one-flavor approximation\footnote{
Throughout the paper, our notation follows Ref.~\cite{boltz}
}, 
\bea 
 \frac{dY_{N_1}}{dz}&=&-\frac{z}{sH(M_1)}
 \left[\left(\frac{Y_{N_1}}{Y_{N_1}^{eq}}-1\right)
 \left(\gamma_{D1}+2\gamma_{h, s}+4\gamma_{h, t}\right)
 +\left(\left[\frac{Y_{N_1}}{Y_{N_1}^{eq}}\right]^2-1\right)
 \left(\gamma_{Z^\prime}+\gamma_{N,t,\Phi}\right)\right], \nonumber \\ 
 \frac{dY_{B-L}}{dz}&=&-\frac{z}{sH(M_1)}\left[\left(\frac{1}{2}
 \frac{Y_{B-L}}{Y_l^{eq}}
 +\epsilon_1\left(\frac{Y_{N_1}}{Y_{N_1}^{eq}}-1\right)\right)
  \gamma_{D_1}   \right.  \nonumber \\
 &+&  \left. 
  \frac{Y_{B-L}}{Y_l^{eq}}
  \left( 2 ( \gamma_N + \gamma_{N, t} + \gamma_{h, t})
 +\frac{Y_{N_1}}{Y_{N_1}^{eq}}\gamma_{h, s}\right)\right], 
\label{eq:be2}
\eea
where $Y_{N_1}$ is the yield (the ratio of the number density to 
 the entropy density $s$) of the (lightest) right-handed neutrino, 
 $Y_{N_1}^{eq}$ is the yield in thermal equilibrium, 
 temperature of the universe is normalized by the mass of the
 right-handed neutrino $z=M_1/T$, 
 $H(M_1)$ is the Hubble parameter at $T=M_1$, 
 $\epsilon_1$ is the CP-asymmetry parameter, 
 and $\gamma$s are the space-time densities of 
 the scatterings in thermal equilibrium.  
Diagrams in Fig.~\ref{gamma} show processes 
 corresponding to different $\gamma$s 
 in the Boltzmann equations, 
 whose explicit forms are listed in Appendix. 
As a good approximation, we neglect masses 
 for all particles involved in the processes, 
 except for the right-handed neutrinos and the $Z'$ boson 
 having the TeV-scale masses. 
The yield of the right-handed neutrino obeys 
 the first equation, while the send equation determines 
 the $B-L$ number created by the out-of-equilibrium decay of 
 the right-handed neutrinos with a non-zero CP-asymmetry parameter. 
In our numerical studies with input parameters given below, 
 we can check that only $\gamma_{D_1}$ and $\gamma_{Z^\prime}$ 
 among the space-time densities have important effects 
 on the final results while the others are negligible.

\begin{figure}[t]
\begin{center}
\includegraphics[scale=1.0]{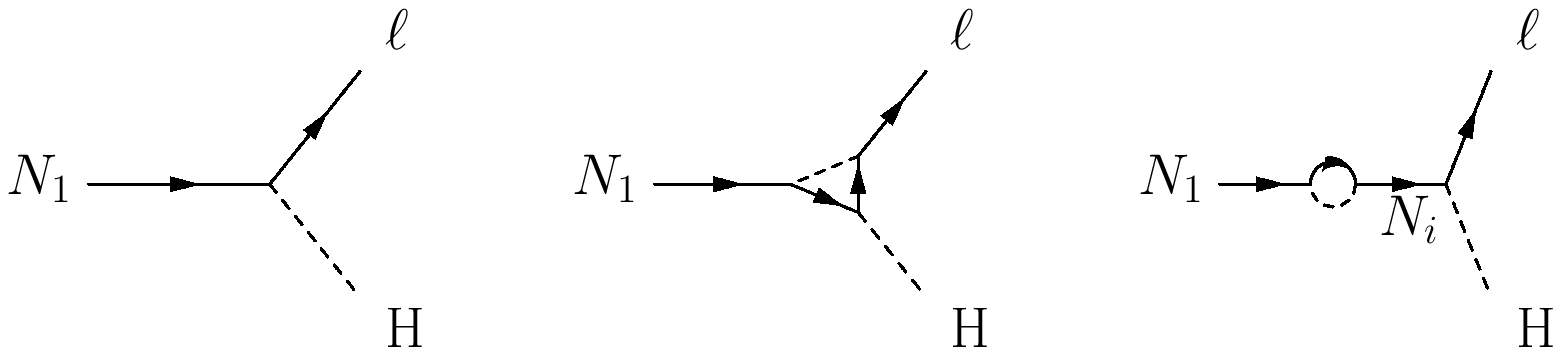}
\caption{
Right-handed neutrino decay at tree and one-loop levels. 
}
\label{fig:cpasym}
\end{center}
\end{figure}

The CP-asymmetry parameter associated 
 with the decay of right-handed neutrino $N_i$ 
 is defined as 
\bea
 \epsilon_i \equiv
 \frac{\sum_j\left[\Gamma\left(N_i \rightarrow \ell_j H \right)
-\Gamma\left(N_i \rightarrow \ell_j^C H^\ast \right)\right]}
      {\sum_j\left[\Gamma\left(N_i \rightarrow \ell_j H\right)
+\Gamma\left(N_i \rightarrow \ell_j^C H^\ast \right)\right]},  
\eea 
 which is generated by the interference between 
 the tree and one-loop diagrams shown in Fig.~\ref{fig:cpasym}.
The general formula is given by \cite{RLG1, RLG} 
\bea 
 \epsilon_i=-\sum_{j \neq i}\frac{M_i}{M_j}
  \frac{\Gamma_j}{M_j}\left(\frac{V_j}{2}+S_j\right)
  \frac{{\rm Im}\left[(y_D y_D^\dagger)_{ij}^2\right]}
  {\left( y_D y_D^\dagger \right)_{ii} 
   \left( y_D y_D^\dagger \right)_{jj}}, 
\label{eq:eps}
\eea 
where 
\begin{equation}
 V_j=2\frac{M_j^2}{M_i^2}\left[\left(1+\frac{M_j^2}{M_i^2}\right)
 \log\left(1+\frac{M_i^2}{M_j^2}\right)  -1\right]
\end{equation}
 corresponds to the vertex correction of 
 the second diagram in Fig.~\ref{fig:cpasym} 
 while 
\bea 
 S_j=\frac{M_j^2 \Delta M_{ij}^2}{\left(\Delta M_{ij}^2\right)^2
 +M_i^2\Gamma_j^2} 
\eea 
 is from the self-energy corrections of the third diagram 
 in Fig.~\ref{fig:cpasym}  
 with the decay width and the mass difference defined as  
\bea 
 \frac{\Gamma_j}{M_j}
  =\frac{\left( y_D y_D^\dagger \right)_{jj}}{8\pi} \; \; {\rm and} \; \;  
 \Delta M_{ij}^2=M_j^2-M_i^2.
\eea 
In the terminology of $K^0-\overline{K^0}$ mixing,
 the first contribution is the so-called direct CP violation
 while the second one the indirect CP violation. 
The indirect CP violation occurs because the mass eigenstates
 and the CP eigenstates are generally different.

When the right-handed neutrinos have a hierarchical mass spectrum 
($M_1 \ll M_{2,3}$), the contributions from 
 $V_j$ and $S_j$ are comparable and 
 the CP-asymmetry parameter is approximately given by 
 \cite{CP asymmetry2}  
\begin{equation}
 \epsilon_1\sim\frac{3}{16\pi}\frac{m_{\nu}M_1}{v^2}\sin\delta
 \sim10^{-6} \left( \frac{m_{\nu}}{0.05{\rm eV}} \right)
 \left( \frac{M_1}{10^{10}{\rm GeV}} \right) \sin\delta, 
\label{eq:cpasym}
\end{equation}
where $m_{\nu}$ is the light neutrino mass eigenvalue, 
 and $\delta$ is the CP-phase and assumed to be of order one. 
The baryon asymmetry of the universe is parameterized as 
\begin{eqnarray}
 Y_B = \kappa\frac{\epsilon_1}{g_*},  
\label{eq:basym}
\end{eqnarray}
where $g_*={\cal O}(100)$ is the number of relativistic degrees 
 of freedom in the early universe, and $\kappa$ is the efficiency factor
 determined by solving the Boltzmann equations, 
 independently of the CP-asymmetry parameter. 
In the leptogenesis scenario without the $Z'$ gauge boson, 
 this factor is roughly estimated as \cite{BBP}
\begin{equation}
 \kappa\sim2\times10^{-2}\left(\frac{0.05{\rm eV}}{m_\nu}\right)^{1.1}. 
\label{eq:effic}
\end{equation}
Using these estimations, we arrive at the conclusion that 
 the Majorana mass $M_1$ has to be larger than $10^{10}$ GeV 
 to give the observed value $Y_B \sim 10^{-10}$. 
However, note that this conclusion is based on 
 the formula of the CP-asymmetry parameter 
 in the case with the hierarchical right-handed neutrino mass spectrum. 
In fact, when two right-handed neutrinos are almost degenerate, 
 the CP-asymmetry parameter can be enhanced.

Now, suppose that two right-handed neutrinos are almost degenerate. 
In this case, it is easy to note that there is a parameter region 
 which can dramatically enhance $S_j$. 
Maximum enhancement occurs for 
 $\Delta M_{ij}^2 \simeq M_i \Gamma_j \ll M_i^2$, 
 so that $S_j \sim M_j/\Gamma_j \gg 1$. 
In this case, the CP-asymmetry parameter is given by  
\begin{equation}
 \epsilon_i \sim  \frac{{\rm Im}\left[(y_D y_D^\dagger)_{ij}^2\right]}
  {\left( y_D y_D^\dagger \right)_{ii}
   \left( y_D y_D^\dagger \right)_{jj}},  
\end{equation}
 which can, in principle, be of order unity.  
The leptogenesis scenario with this enhancement 
 of the CP-asymmetry parameter is called the
 resonant leptogenesis \cite{RLG}. 
This enhancement is crucial to realize the observed baryon asymmetry 
 when the right-handed neutrino mass is significantly smaller 
 than $10^{10}$ GeV, such as the TeV scale 
which is  of our main concern in this paper.

\subsection*{Analysis of the Boltzmann equations in one-flavor approximation}

Now we analyze the Boltzmann equations in Eq.~(\ref{eq:be2}) 
 to see how much baryon asymmetry can be generated 
 in the $B-L$ model at the TeV scale. 
In order to understand the response 
 between the model-parameters involved in this analysis 
 and the resultant baryon asymmetry, 
 we first consider a  model in one-flavor approximation 
 with the parameterization of the decay width as 
\bea 
  \Gamma_1 = \frac{y_D^2}{8 \pi} M_1   
\eea
 with a real free parameter $y_D$, 
 while the other parameters are fixed as follows: 
\bea 
 \epsilon_1=0.01, \; \alpha_{B-L}=0.006, \; 
 m_{Z'}=3 \; {\rm TeV}, \; M_1 = 2 \; {\rm TeV}.  
\eea 
Then, we numerically solve the Boltzmann equations 
 with the boundary conditions 
\bea 
  Y_{N_1}(0) = Y_{N_1}^{eq}(0), \; Y_{B-L}(0) =0.    
\eea
The lepton asymmetry generated by the right-handed neutrino decays 
 is converted into the baryon asymmetry via 
 the sphaleron process while the process is in thermal equilibrium. 
In our analysis throughout the paper, we evaluate 
 the resultant baryon number at the freeze-out temperature 
 of the sphaleron process, $T_{sph} \simeq 150$ GeV \cite{Tsph}, 
 where the conversion of the lepton number to the baryon number 
 is terminated: 
\bea 
  Y_B = \frac{28}{79} Y_{B-L}(z_{sph}), 
\eea  
where $Y_{B-L}(z_{sph})$ is the numerical solution 
 of the Boltzmann equations at $z_{sph}=M_1/T_{sph}$.

The resultant baryon asymmetry is depicted in Fig.~3 
 as a function of $y_D$ (solid line). 
For comparison, results for the cases 
 with $\alpha_{B-L}=0$ (dotted blue line) 
 and with $\gamma_{N,t,\Phi}=0$ as well as $\alpha_{B-L}=0$ 
 (dashed green line) are also shown. 
For a small $y_D^2 \lesssim 10^{-10.5}$, we can see 
 that the generation of the baryon asymmetry is 
 suppressed in the presence of the Z' boson and 
 $\gamma_{N,t,\Phi}$ processes. 
Although the suppression by the $Z'$ boson effect dominates, 
 the $\gamma_{N,t,\Phi}$ process mediated by the Majorana 
 Yukawa coupling ($y_N$) causes a dramatic reduction 
 in generating the baryon asymmetry even in the absence 
 of the $Z'$ boson effect. 
In the region, $Y_B$ is growing as $y_D$, and  
 a larger $y_D$ generates a larger baryon asymmetry 
 against the $Z'$ boson and Majorana  Yukawa coupling effects. 
On the other hand, for $y_D^2 \gtrsim 10^{-10}$, 
 the effect by Dirac Yukawa coupling dominates 
 over the $Z'$ boson and Majorana Yukawa coupling effects, 
 and all lines become well-overlapping. 
In this region, however, $Y_B$ is suppressed 
 by the washing-out process via the inverse-decay process. 
In the dashed (green) line, 
 $Y_B$ becomes smaller as $y_D^2$ is lowered 
 for $y_D^2 \lesssim 10^{-15}$ GeV, 
 nevertheless $\gamma_{N,t,\Phi}=0$ and $\alpha_{B-L}=0$. 
This is because the generation of lepton number is 
 too slow with such a small Dirac Yukawa coupling, and 
 the sphaleron process freezes out before the completion 
 of the whole lepton number generation.

\begin{figure}[t]
\begin{center}
{\includegraphics[scale=0.8]{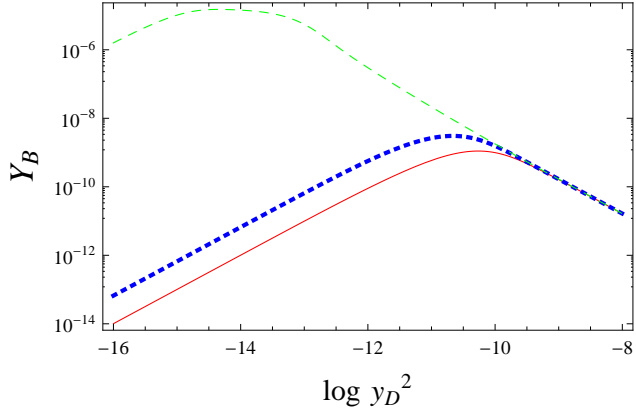}}
\caption{
The baryon asymmetry generated in the minimal $B-L$ model 
 (solid red line) as a function of the Dirac Yukawa coupling. 
The dotted (blue) line corresponds to the result of 
 baryon asymmetry in the absence of the $B-L$ gauge interaction, 
 while the dashed (green) line is for the case 
 with $\gamma_{N,t,\Phi}=0$ as well as $\alpha_{B-L}=0$. 
}
\label{fig3}
\end{center}
\end{figure}

\begin{figure}[t]\begin{center}
{\includegraphics[scale=0.8]{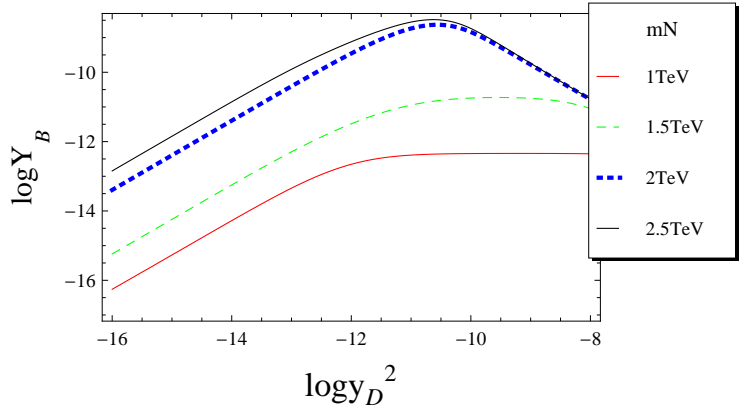}}
\caption{
The baryon asymmetry as a function of Yukawa coupling 
 for different values of the right-handed neutrino mass, 
 $M_1=1$ TeV (solid red line), $M_1=1.5 $TeV (dashed green line) 
 , $M_1=2$ TeV (dotted blue line) and 
 $M_1=2.5$ TeV (solid black line). 
}
\label{figyukawa}
\end{center}
\end{figure}

Fig.~\ref{figyukawa} shows the results for different values of $M_1$ 
 as a function of $y_D^2$ while the other parameters are kept the same. 
We can see, for $M_1 > 1.5 $ TeV, a similar behavior 
 to the result shown in Fig.~3. 
For a relatively small $M_1 \lesssim 1.5$ TeV, 
 the resultant $Y_B$ becomes almost independent of $y_D^2$ 
 even for a larger $Y_D$. 
This is because the freeze-out of the sphaleron process occurs  
 and thus the conversion of the lepton number to the baryon number 
 is terminated before the washing-out process becomes effective.

\begin{figure}[t]\begin{center}
{\includegraphics[scale=0.8]{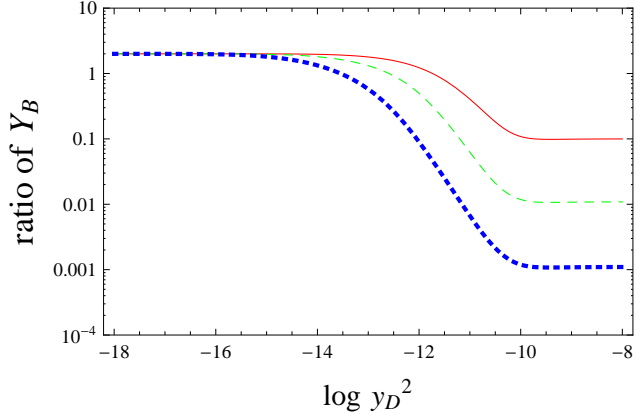}}
\caption{
The ratio $Y_{B-L}^{\rm 2-flavor}/Y_{B-L}^{\rm 1-flavor}$
 as the Dirac Yukawa coupling squared $y_D^2$,  
 for different choices of 
 $\frac{\Gamma_2}{\Gamma_1}=10$ (red solid line) 
 $\frac{\Gamma_2}{\Gamma_1}=100$ (green dashed line) 
 and  $\frac{\Gamma_2}{\Gamma_1}=1000$ (blue dotted line).   
}
\label{fig2gene}
\end{center}
\end{figure}

\subsection*{2-flavor analysis}

We have so far investigated the leptogenesis 
 by solving the Boltzmann equations with one-flavor 
 right-handed neutrino. 
This treatment is justified when the right-handed 
 neutrinos have a hierarchical mass spectrum. 
In the resonant leptogenesis, two right-handed neutrinos 
 are degenerated in mass and thus, it is non-trivial 
 whether one-flavor analysis is actually a good 
 approximation or not. 
Here, we generalize our analysis to the 2-flavor case 
 and clarify when the 1-flavor analysis is justified.

As we have discussed in the previous section, 
 the CP-asymmetry parameter $\epsilon_i$ is maximally enhanced 
 when two right-handed neutrinos are almost degenerate 
 and their mass squared difference is 
 $\Delta M^2_{ij} \simeq M_i \Gamma_j$. 
Thus, in our analysis for the two-flavor case, 
 we set the mass difference as 
 $\Delta M_{12}^2 = M_1 \Gamma_2$. 
Note that since $M_{1,2} \gg \Gamma_{1,2}$ and thus $M_1 \simeq M_2$, 
 the case with $\Delta M_{12}^2 = M_1 \Gamma_1$ 
 is essentially the same as the case 
 with the exchange $1 \leftrightarrow 2$. 
The CP-asymmetry parameters are given by 
\bea 
 \epsilon_1 &\simeq& -\frac{1}{2} 
 \frac{{\rm Im}\left[(y_D y_D^\dagger)_{12}^2\right]}
 {\left( y_D y_D^\dagger \right)_{11} 
  \left( y_D y_D^\dagger \right)_{22}}, \nonumber\\
 \epsilon_2 &\simeq& 
   \epsilon_1 \times \frac{2\Gamma_1\Gamma_2}{\Gamma_1^2+\Gamma_2^2},  
\eea 
where we have used the relations 
 $(y_D y_D^\dagger)_{12} = (y_D y_D^\dagger)_{21}^*$ 
 and $\Delta M_{12}^2=-\Delta M_{21}^2$.

We consider the following three cases;
\begin{enumerate}
\item $\Gamma_1\gg\Gamma_2$ \\
      The CP asymmetry parameter 
      $\epsilon_2 \simeq 2 \epsilon_1 {\Gamma_2}/{\Gamma_1} \ll
      \epsilon_1$, and the baryon asymmetry is generated dominantly 
      by the $N_1$ decay. 
      In addition, the washing-out process by the inverse decay 
      of $N_2$ is also negligible to that by $N_1$. 
      Therefore, analysis with only one-flavor right-handed neutrino 
      $N_1$ is sufficient in evaluating the resultant baryon asymmetry. 

\item $\Gamma_1\sim\Gamma_2$ \\
      Clearly, two right-handed neutrinos are almost identical, 
      so that one-flavor analysis is sufficient, but the resultant 
      baryon asymmetry should be twice of that obtained 
      in one-flavor case. 
 
\item $\Gamma_1\ll\Gamma_2$ \\
      The CP asymmetry parameter 
      $\epsilon_2 \simeq 2 \epsilon_1 \frac{\Gamma_1}{\Gamma_2} 
      \ll \epsilon_1$ and hence the generation of baryon asymmetry 
      by $N_2$ decays are negligible. 
      However, the washing-out effect by the inverse decay of $N_2$ 
      can be efficient and the generated baryon asymmetry can be 
      drastically reduced, depending on the value of $\Gamma_2$. 
\end{enumerate} 
Therefore, for the cases 1 and 2, 
 one-flavor analysis is sufficient to evaluate generated 
 baryon asymmetries, while the case 3 is non-trivial 
 and we need to solve the Boltzmann equations with two-flavor 
 right-handed neutrinos.

The Boltzmann equations with two flavor right-handed neutrinos 
 are given by 
\bea 
\frac{dY_{N_1}}{dz}&=&-\frac{z}{sH(M_1)}\left[
\left(\frac{Y_{N_1}}{Y_{N_1}^{eq}}-1\right)\gamma_{D_1}
+\left(\left(\frac{Y_{N_1}}{Y_{N_1}^{eq}}\right)^2-1\right)\gamma_{Z'}
\right], \nonumber \\ 
\frac{dY_{N_2}}{dz}&=&-\frac{z}{sH(M_1)}\left[
\left(\frac{Y_{N_2}}{Y_{N_2}^{eq}}-1\right)\gamma_{D_2}
+\left(\left(\frac{Y_{N_2}}{Y_{N_2}^{eq}}\right)^2-1\right)\gamma_{Z'}
\right], \nonumber \\  
\frac{dY_{B-L}}{dz}&=&-\frac{z}{sH(M_1)}\left[{\sum^2_{j=1}}
\left(\frac{1}{2}\frac{Y_{B-L}}{Y_l^{eq}}
+\epsilon_j\left(\frac{Y_{N_j}}{Y_{N_j}^{eq}}-1\right)\right)\gamma_{D_j}
\right].  
\label{2-f BE}
\eea 
Here we have omitted $\gamma$s mediated by the
 Higgs boson and the right-handed neutrinos, 
 because these effects are, in fact, negligible. 
However, all processes in Fig.~1 are taken into account 
 in our numerical analysis. 
With the initial conditions, $Y_{N_i}(0)=Y_{N_i}^{eq}(0)$ $(i=1,2)$ 
 and $Y_{B-L}(0)=0$, we solve these equations and evaluate 
 the baryon number at $T_{sph}=150$ GeV. 
In order to compare results to those in the one-flavor case, 
 we parameterize the decay width as $\Gamma_1 =y_D^2 M_1/(8 \pi)$ 
 corresponding to the one-flavor case. 
For fixed values of $\Gamma_2/\Gamma_1$,  
 we show the ratio of the baryon asymmetry to 
 the one obtained in the one-flavor analysis,  
 $Y_{B-L}^{\rm 2-flavor}/Y_{B-L}^{\rm 1-flavor}$, 
 in Fig.~\ref{fig2gene}. 
For a small $y_D$, we can see that the one-flavor analysis is 
 a good approximation. 
On the other hand, the washing-out process by 
 the inverse-decay of $N_2$ is very effective 
 for a large $y_D$, and the baryon asymmetry 
 is very much suppressed than the result obtained 
 in the one-flavor analysis. 
As is expected, the baryon asymmetry is more suppressed 
 as the ratio $\Gamma_2/\Gamma_1$ becomes larger. 
Therefore, in order to obtain the correct result 
 for baryon asymmetry via the resonant leptogenesis, 
 analysis with two (or more) flavors can be essential 
 in the general case, especially, when $\epsilon_i \ll \epsilon_j$ 
 but $\Gamma_i \gg \Gamma_j$ for two almost degenerate 
 right-handed neutrinos, $N_i$ and $N_j$.

\section{Neutrino oscillation data and resonant leptogenesis} 
In the previous section, we have analyzed the resonant leptogenesis 
 in the minimal $B-L$ model 
 and investigated the response of the resultant baryon asymmetry  
 to model-parameters such as the Dirac Yukawa couplings and the
 right-handed neutrino masses. 
It is clearly more interesting to consider a realistic model 
 (in other words, a realistic parameterization)  
 which can account for the observed neutrino oscillation phenomena. 
For this purpose, we consider a strategy first proposed 
 in \cite{BO} in this section. 
In our analysis, we adopt the current neutrino oscillation data 
 in 2-$\sigma$ range \cite{oscidata}:
\bea 
 &7.25\times10^{-5} < \Delta m_{12}^2 ({\rm eV}^2) < 8.11 \times10^{-5}, 
  \nonumber\\
 &2.18\times10^{-3} <|\Delta m_{13}^2|({\rm eV}^2) < 2.64 \times10^{-3}, 
  \nonumber\\
 &0.27<\sin^2\theta_{12}<0.35, \nonumber\\
 &0.39<\sin^2\theta_{23}<0.63, \nonumber\\
 &\sin^2\theta_{13}\leq0.040, 
\eea
for the standard parameterization of the mixing matrix, 
\begin{eqnarray}
U_{PMNS}=\left(
\begin{array}{ccc}
 {\scriptstyle \cos\theta_{12}\cos\theta_{13}} 
 & {\scriptstyle \sin\theta_{12}\cos\theta_{13}} 
 & {\scriptstyle\sin\theta_{13}e^{-i\delta}} \\
 {\scriptstyle-\sin\theta_{12}
  -\cos\theta_{12}\sin\theta_{23}\cos\theta_{13}e^{i\delta}}
 & {\scriptstyle\cos\theta_{12}\cos\theta_{23}
  -\sin\theta_{12}\sin\theta_{23}\sin\theta_{13}e^{i\delta}} 
 & {\scriptstyle\sin\theta_{23}\cos\theta_{13}} \\
 {\scriptstyle\sin\theta_{12}\sin\theta_{23}
  -\cos\theta_{12}\cos\theta_{23}\sin\theta_{13}e^{i\delta}} 
 & {\scriptstyle-\cos\theta_{12}\cos\theta_{23}
  -\sin\theta_{12}\cos\theta_{23}\sin\theta_{13}e^{i\delta}} 
 & {\scriptstyle\cos\theta_{23}\cos\theta_{13}} \\
\end{array}
\right)\nonumber\\
\times {\rm diag}(e^{i\frac{\alpha_1}{2}},e^{i\frac{\alpha_2}{2}},1),  
\label{PMNS}
\end{eqnarray}
with the Dirac phase $\delta$ and the Majorana phases $\alpha_i$.

We consider the so-called minimal seesaw \cite{FGY} 
 in the context of the minimal $B-L$ model, and assume 
 that only two right-handed neutrinos are relevant 
 for the neutrino oscillation phenomena and leptogenesis. 
The third right-handed neutrino is assumed to decouple 
 from the neutrino oscillation phenomena by some reason. 
A simple idea is to introduce a discrete $Z_2$ symmetry 
 under which the third right-handed neutrino is assigned 
 to be odd while all the other particles in the $B-L$ model even. 
In the context of the minimal $B-L$ model with 
 the $Z_2$ symmetry, it has been shown \cite{OS} that 
 the third right-handed neutrino can be a suitable candidate 
 for the cold dark matter with the relic density consistent 
 with observations.

In the minimal seesaw model, 
 we parameterize the $2 \times 3$ Dirac neutrino mass matrix, 
 without loss of generality, as 
\bea 
m_D=\left(
 \begin{array}{ccc}
 a_1e^{i\phi_1} & a_2e^{i\phi_2} & a_3e^{i\phi_3} \\
 a_4 & a_5 & a_6 \\
\end{array}
\right),  
\label{dirac}
\eea 
where $a_i$ and $\phi_j$ are real parameters, and 
 we have worked in the basis 
 where both the charged lepton mass matrix and the
 right-handed neutrino mass matrix are diagonalized 
 with real and positive eigenvalues. 
We parameterize the Majorana mass matrix of the two right-handed neutrinos as 
\bea 
 M_N=\left(
\begin{array}{cc}
 M_1 & 0 \\
 0 & M_1(1+r) \\
\end{array}\right),   
\end{eqnarray}
where the parameter $r$ should be very small, 
 for example, $ r \sim \Gamma_1/M_1$ or $\Gamma_2/M_1$ 
 in order to realize the enhancement of 
 the CP-asymmetry parameter. 
Although $r$ is crucial for the resonant leptogenesis, 
 such a small $r$ is negligible in fitting 
 for the neutrino oscillation data.

\begin{figure}[t]
\begin{tabular}{cc}
\begin{minipage}{0.5\hsize}
\begin{center}
{\includegraphics[scale=.8]{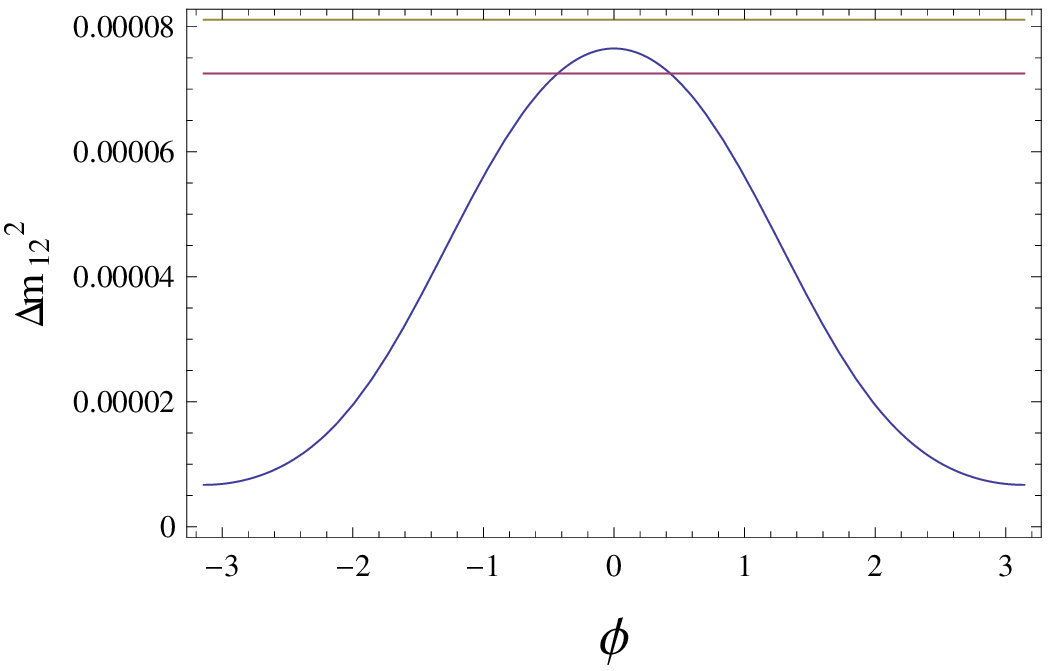}}
\end{center}
\end{minipage}
\begin{minipage}{0.5\hsize}
\begin{center}
{\includegraphics[scale=.8]{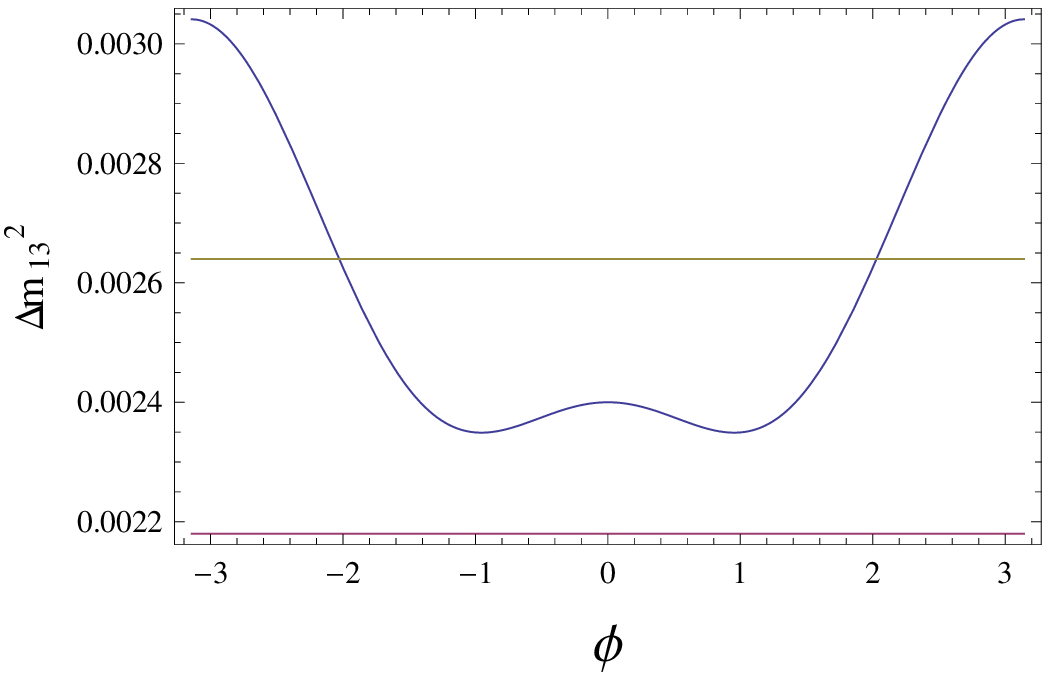}}
\end{center}
\end{minipage}
\end{tabular}
\caption{
 The neutrino oscillation parameters, 
 $\Delta m_{12}$ (left panel) and 
 $\Delta m_{13}$ (right panel), 
 as a function of the CP-phase $\phi_3$. 
The observed data in 2-$\sigma$ range 
 are indicated by two horizontal lines. 
}
\label{fig:massNH}
\end{figure}
\begin{figure}[t]
\begin{tabular}{cc}
\begin{minipage}{0.5\hsize}
\begin{center}
{\includegraphics[scale=.8]{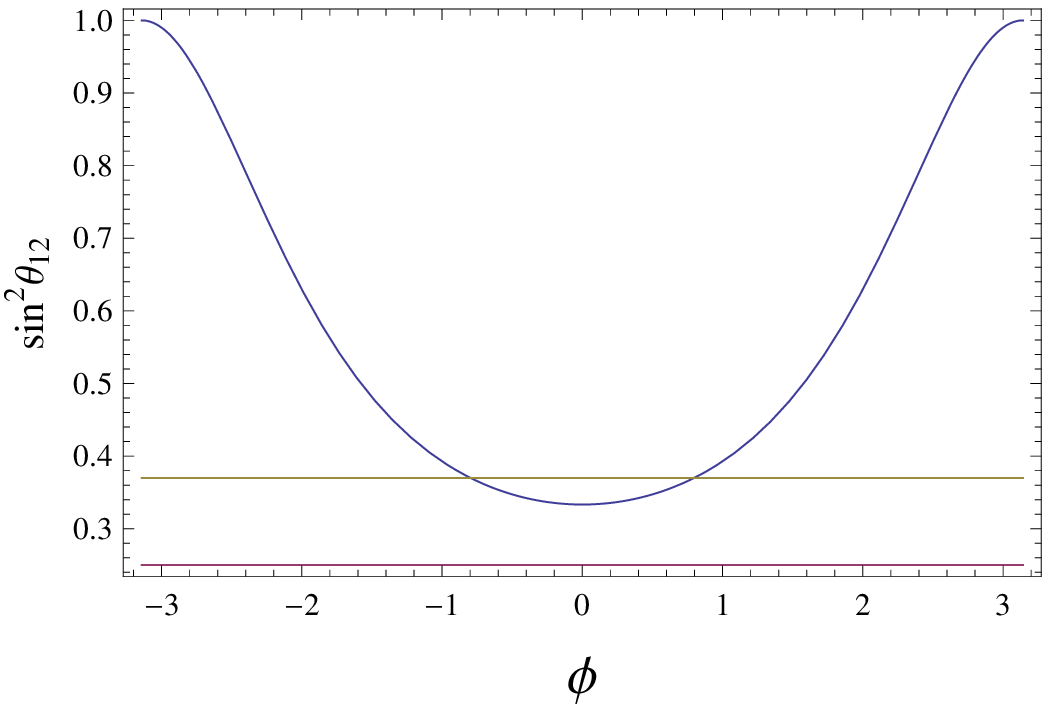}}
\end{center}
\end{minipage}
\begin{minipage}{0.5\hsize}
\begin{center}
{\includegraphics[scale=.8]{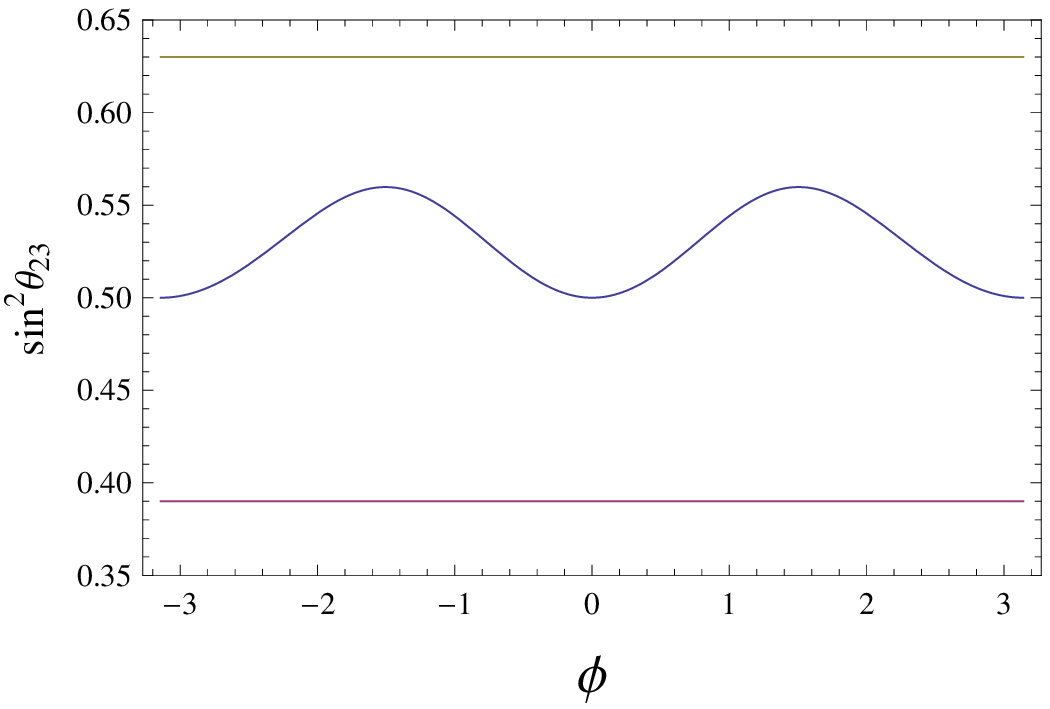}}
\end{center}
\end{minipage}
\end{tabular}
\caption{
The neutrino oscillation parameters, 
 $\sin^2\theta_{12}$ (left panel) and 
 $\sin^2\theta_{23}$ (right panel),  
 as a function of the CP-phase $\phi_3$. 
The observed data in 2-$\sigma$ range 
 are indicated by two horizontal lines. 
}
\label{fig:osc1NH}
\end{figure}
\begin{figure}[t]\begin{center}
{\includegraphics[scale=0.8]{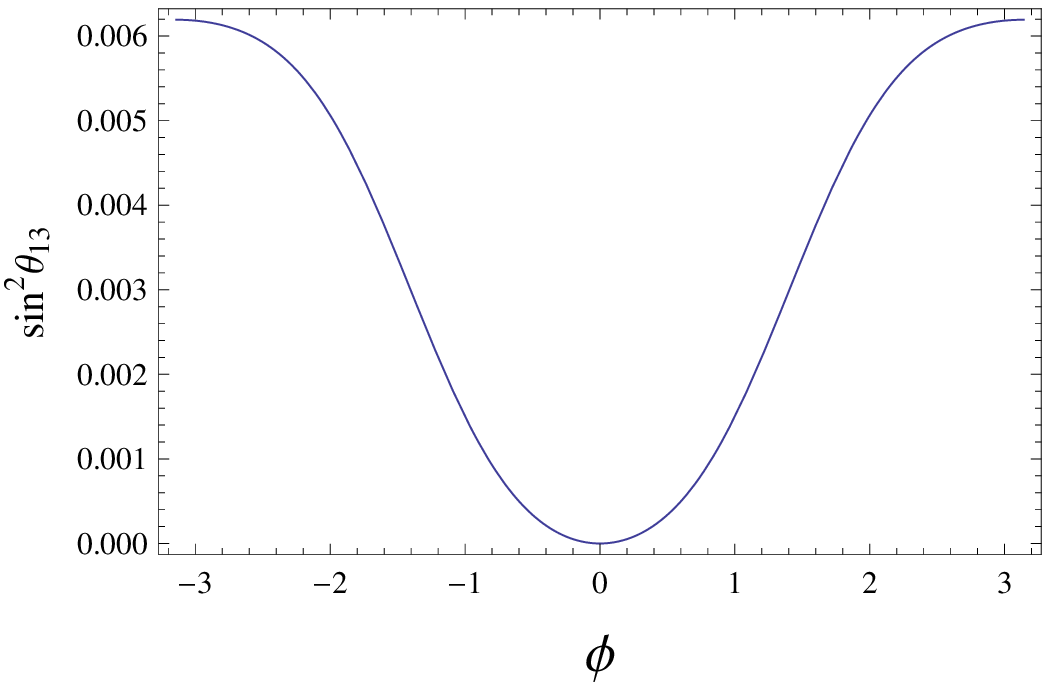}}
\caption{
 $\sin^2\theta_{13}$ as a function of CP phase.
}
\label{fig:osc2NH}
\end{center}
\end{figure}

\begin{figure}[t]\begin{center}
{\includegraphics[scale=0.8]{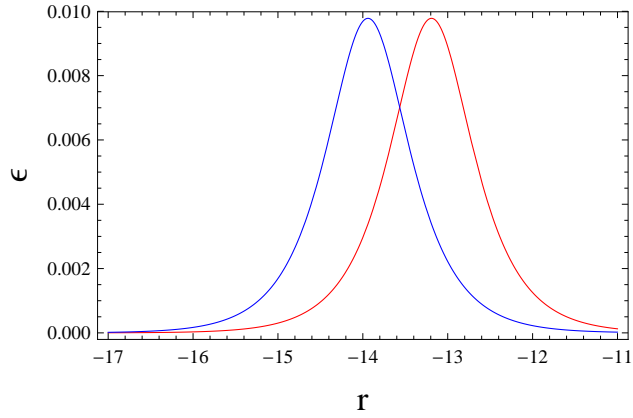}}
\caption{
The CP asymmetry parameters, $\epsilon_1$ (right) 
 and $\epsilon_2$ (left), as a function of $r$. 
}
\label{fig:NHeps}
\end{center}
\end{figure}

\begin{figure}[t]\begin{center}
{\includegraphics[scale=0.8]{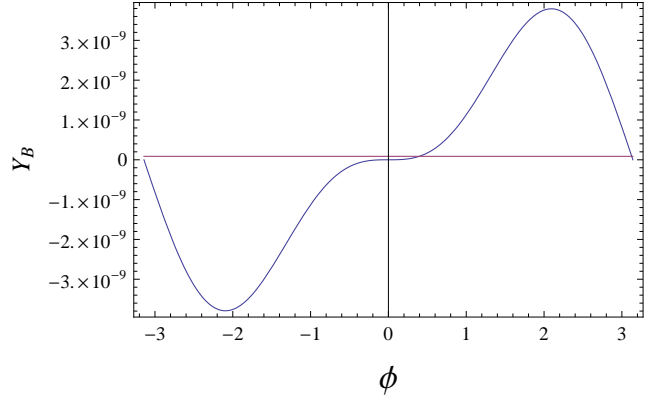}}
\caption{
The resultant baryon asymmetry in the universe 
 as a function of the CP-phase $\phi_3$.
The observed value $Y_B=0.87 \times 10^{-10}$ 
 is depicted as the horizontal line. 
}
\label{fig:NHyb}
\end{center}
\end{figure}

For simplicity, we fix $\phi_1=\phi_2=0$ in our analysis 
 and introduce  an ansatz \cite{BO} that the light neutrino mass 
 matrix after the seesaw mechanism \cite{seesaw}, 
\bea 
 m_\nu= m_D^T M_N^{-1} m_D 
      \simeq \frac{1}{M_1} m_D^T m_D , 
\eea
 is diagonalized by the so-called tri-bimaximal mixing matrix 
 \cite{TBmatrix},  
\bea 
 U_{TB}=\left(
\begin{array}{ccc}
 \sqrt{\frac{2}{3}} & \sqrt{\frac{1}{3}} & 0 \\
 -\sqrt{\frac{1}{6}} & \sqrt{\frac{1}{3}} & \sqrt{\frac{1}{2}} \\
 -\sqrt{\frac{1}{6}} & \sqrt{\frac{1}{3}} & -\sqrt{\frac{1}{2}} \\
\end{array}
\right), 
\eea 
 in the CP invariant case ($\phi_3=0$). 
As is well-known, this tri-bimaximal mixing matrix gives 
 almost the best fit in the oscillation data.

Note that in the minimal seesaw, the rank of 
 the light neutrino mass matrix is two 
 and the lightest mass eigenvalue is 0. 
In the following we consider two cases for 
 the light neutrino mass spectrum, namely, 
 the normal hierarchical (NH) case and 
 inverted hierarchical (IH) case. 
Let us first consider the NH case, where we have 
\begin{eqnarray}
D_\nu^{NH}={\rm diag} \left( 
  0, m_2^{NH}, m_3^{NH} \right)  
\end{eqnarray}
with $m_2^{NH} = \sqrt{\Delta m_{12}^2}$ and 
 $m_3^{NH} = \sqrt{|\Delta m_{13}^2|}$.  
According to our ansatz, we first find 
 a solution to $m_\nu = U_{TB} D_\nu^{NH} U_{TB}^T$ 
 in the CP-invariant case. 
Among several solutions, we choose, as an example, 
\begin{eqnarray}
 a_1&=&a_2=a_3=\sqrt{\frac{M_1 m_2^{NH}}{3}},\nonumber\\
 a_4&=&0,\nonumber\\
 a_5&=&-a_6=\sqrt{\frac{M_1 m_3^{NH}}{2}}. 
\label{NH}
\end{eqnarray}
For the input values, we use 
 $m_2^{NH}=8.75 \times 10^{-3}$ eV 
 and $m_3^{NH}= 4.90 \times 10^{-2}$ eV. 
In this section, we fix other parameters 
 as $\alpha_{B-L}=0.006$, $m_{Z'}=3$ TeV and $M_1=2$ TeV.

Now we turn the CP-phase $\phi_3$ on. 
With fixed $a_i$ and $M_1$, the light neutrino mass matrix 
 is given as a function of the single parameter $\phi_3$ 
 \cite{BO} (in the approximation with $r=0$). 
In Figs.~6$-$8, the neutrino oscillation parameters are depicted  
 as a function of the CP-phase $\phi_3$. 
As we expect, the outputs of the oscillation parameters 
 deviate from the values at $\phi_3=0$ as the CP-phase is changed, 
 and eventually some of outputs are found to be outside 
 of 2-$\sigma$ range of the experimental data. 
We find the bound on the CP-phase as $|\phi_3| \lesssim 0.5$.

For the resonant leptogenesis, both $\phi_3 \neq 0$ and $r \neq 0$ 
 are crucial. 
Fig.~9 shows the CP-asymmetry parameters 
 ($\epsilon_1$ and $\epsilon_2$) as a function of $r$ 
 with $\phi_3 =0.5$ for example. 
For the right (left) curve corresponding to $\epsilon_1$
 ($\epsilon_2$), a peak appears around $r=10^{-13}$ 
 ($r=10^{-14}$). 
Choosing $r=10^{-14}$ for example, the CP-asymmetry parameters  
  are also given as a function of only $\phi_3$. 
Therefore, we have correlations between neutrino oscillation 
 parameters and the baryon asymmetry generated by the resonant 
 leptogenesis through the CP-phase $\phi_3$ \cite{BO}. 
Interestingly, the amount of the generated baryon asymmetry becomes larger 
 as $\phi_3$ goes away from zero, 
 while a large displacement of $\phi_3$ from zero 
 results in the oscillation parameters inconsistent with 
 the experimental data.

Numerical solution of the Boltzmann equations with two flavors 
 in Eq.~(\ref{2-f BE}) is shown Fig.~10 
 as a function of $\phi_3$. 
We find that the observed baryon asymmetry 
 $Y_B=0.87\times 10^{-10}$ in the present universe 
 is obtained for $\phi_3 = 0.35$, for which 
 the neutrino oscillation parameters are fixed as  
\begin{eqnarray}
 \Delta m_{12}^2 ({\rm eV}^2) &=& 7.39 \times10^{-5}, 
\nonumber\\
 \Delta m_{13}^2 ({\rm eV}^2) &=& 2.39 \times10^{-3}, 
\nonumber\\
 \sin^2\theta_{12}&=&0.34, 
\nonumber\\
 \sin^2\theta_{23}&=&0.51,  
\nonumber\\
 \sin^2\theta_{13}&=&0.00016. 
\end{eqnarray}
They are all consistent with observations. 
Although a non-vanishing $\sin^2\theta_{13}$ is predicted, 
 it is quite small, far below the current upper bound.

\begin{figure}[t]
\begin{tabular}{cc}
\begin{minipage}{0.5\hsize}
\begin{center}
{\includegraphics[scale=.8]{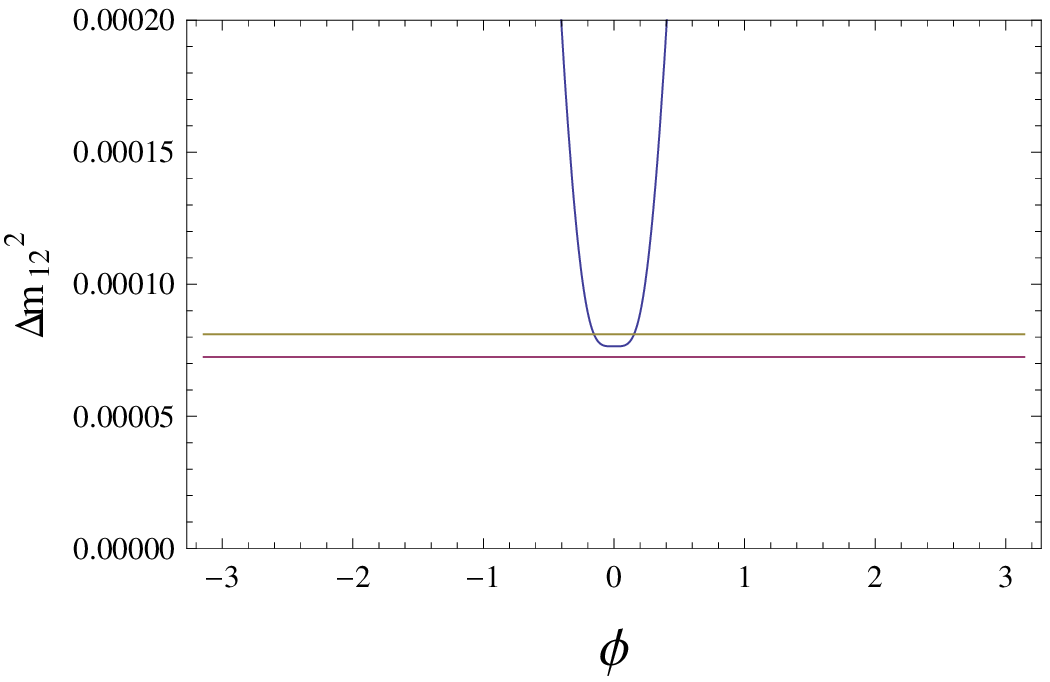}}
\end{center}
\end{minipage}
\begin{minipage}{0.5\hsize}
\begin{center}
{\includegraphics[scale=.8]{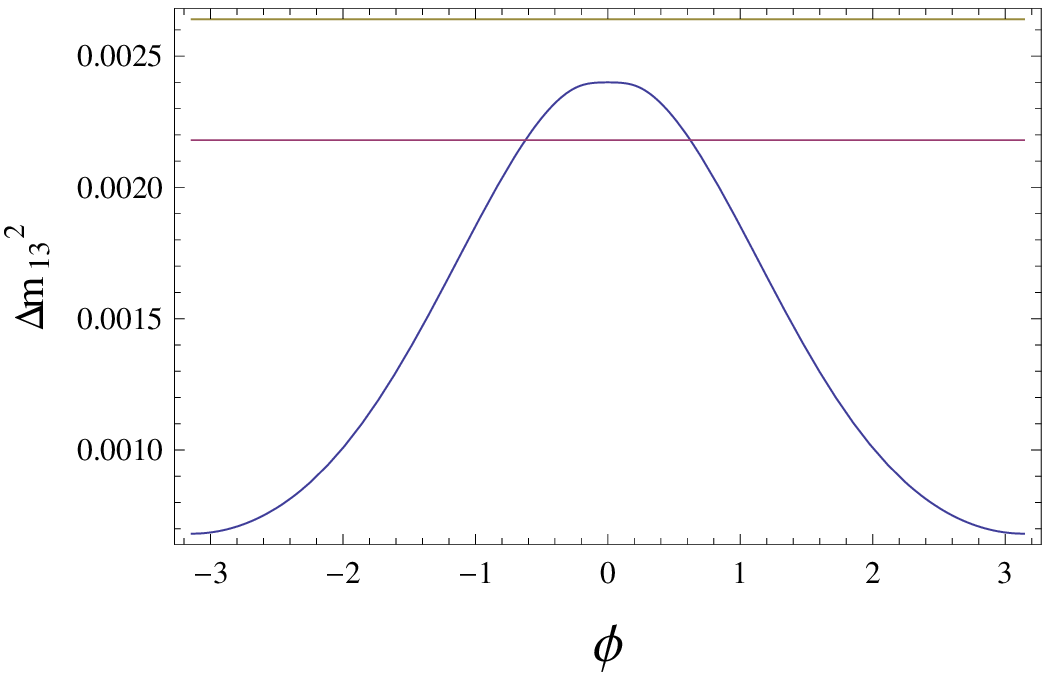}}
\end{center}
\end{minipage}
\end{tabular}
\caption{
The neutrino oscillation parameters, 
 $\Delta m_{12}$ (left panel) and 
 $\Delta m_{13}$ (right panel), 
 as a function of CP-phase $\phi_3$. 
The observed data in 2-$\sigma$ range 
 are indicated by two horizontal lines. 
}
\label{fig:massIH}
\end{figure}

\begin{figure}[t]
\begin{tabular}{cc}
\begin{minipage}{0.5\hsize}
\begin{center}
{\includegraphics[scale=.8]{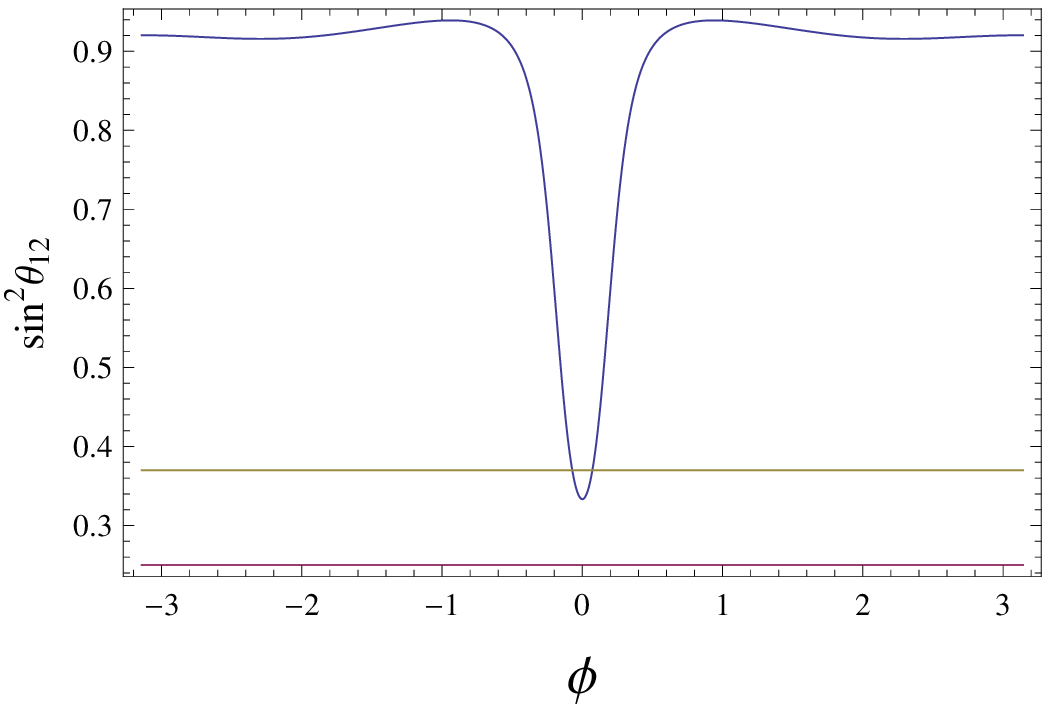}}
\end{center}
\end{minipage}
\begin{minipage}{0.5\hsize}
\begin{center}
{\includegraphics[scale=.8]{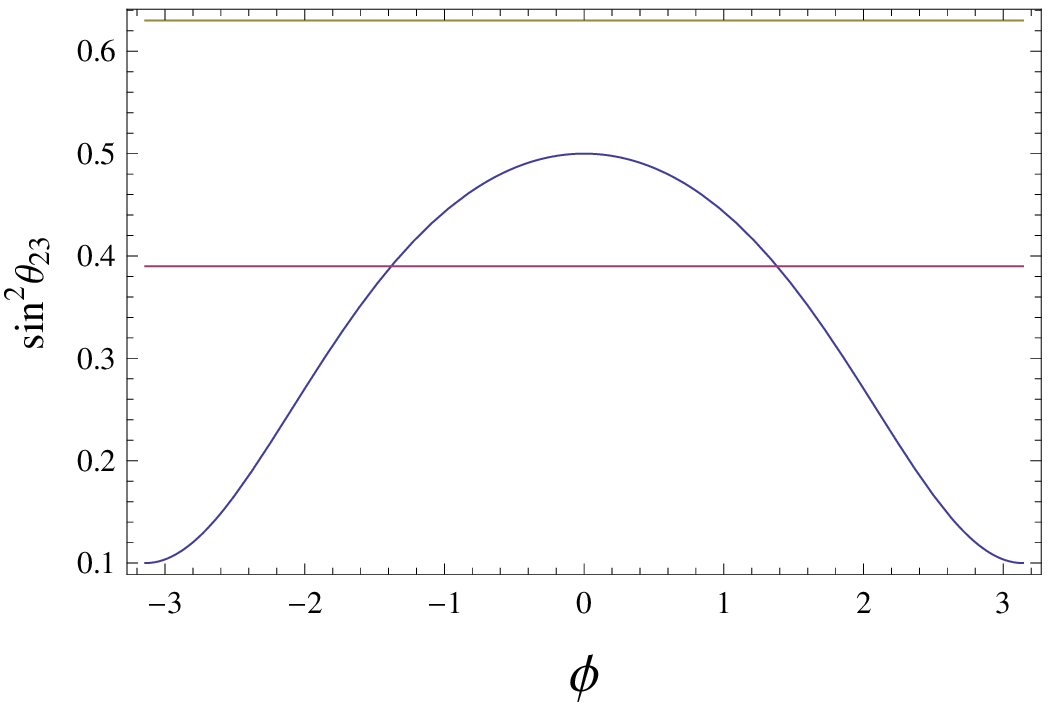}}
\end{center}
\end{minipage}
\end{tabular}
\caption{
The neutrino oscillation parameters, 
 $\sin^2\theta_{12}$ (left panel) and 
 $\sin^2\theta_{23}$ (right panel), 
 as a function of CP-phase $\phi_3$. 
The observed data in 2-$\sigma$ range 
 are indicated by two horizontal lines. 
}
\label{fig:osc1IH}
\end{figure}

\begin{figure}[t]\begin{center}
{\includegraphics[scale=0.8]{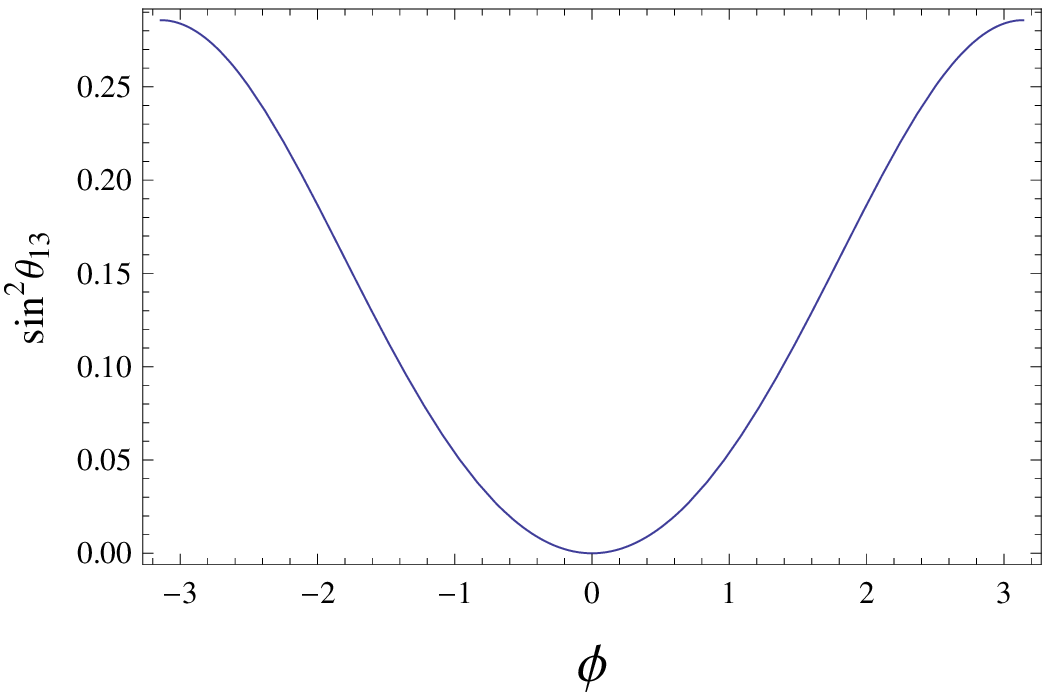}}
\caption{
$\sin^2\theta_{13}$ as a function of the CP-phase $\phi_3$. 
}
\label{fig:osc2IH}
\end{center}
\end{figure}

Next we consider the IH case, where the light neutrino mass 
 matrix is diagonalized as 
\begin{eqnarray}
 D_\nu^{IH}= {\rm diag} \left(
  m_1^{IH}, m_2^{IH},  0 \right), 
\end{eqnarray}
 with $m_1^{IH}=\sqrt{|\Delta m_{13}^2|}$ and 
      $m_2^{IH}=\sqrt{\Delta m_{12}^2+|\Delta m_{13}^2|}$.  
In the CP invariant case, we choose 
 a solution to $m_\nu = U_{TB} D_\nu^{IH} U_{TB}^T$ as  
\begin{eqnarray}
 a_1&=&a_2=a_3=\sqrt{\frac{M_1 m_2^{IH}}{3}},\nonumber\\
 a_4&=&\sqrt{\frac{2M_1 m_1^{IH}}{3}}, \nonumber\\
 a_5&=&a_6=-\sqrt{\frac{M_0m_{1, IH}}{6}}. 
\label{IH}
\end{eqnarray}
For the input values of 
 $m_1^{IH} =4.90 \times10^{-2}$ eV and 
 $m_2^{IH} = 4.98 \times 10^{-2}$ eV, 
 the neutrino oscillation parameters are depicted 
 in Figs.~\ref{fig:massIH}$-$\ref{fig:osc2IH} 
 as a function of $\phi_3$. 
A CP-phase $|\phi_3| \lesssim 0.1$ results 
 in the outputs of the neutrino oscillation parameters 
 consistent with the experimental data in 2-$\sigma$ range.


\begin{figure}[t]\begin{center}
{\includegraphics[scale=0.8]{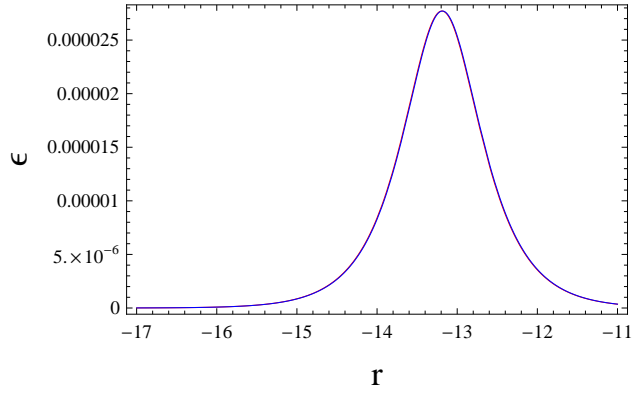}}
\caption{
The CP asymmetry parameters, 
 $\epsilon_1$ and $\epsilon_2$
 as a function of $r$. 
Two curves are well-overlapped. 
}
\label{fig:IHeps}
\end{center}
\end{figure}

\begin{figure}[t]\begin{center}
{\includegraphics[scale=0.8]{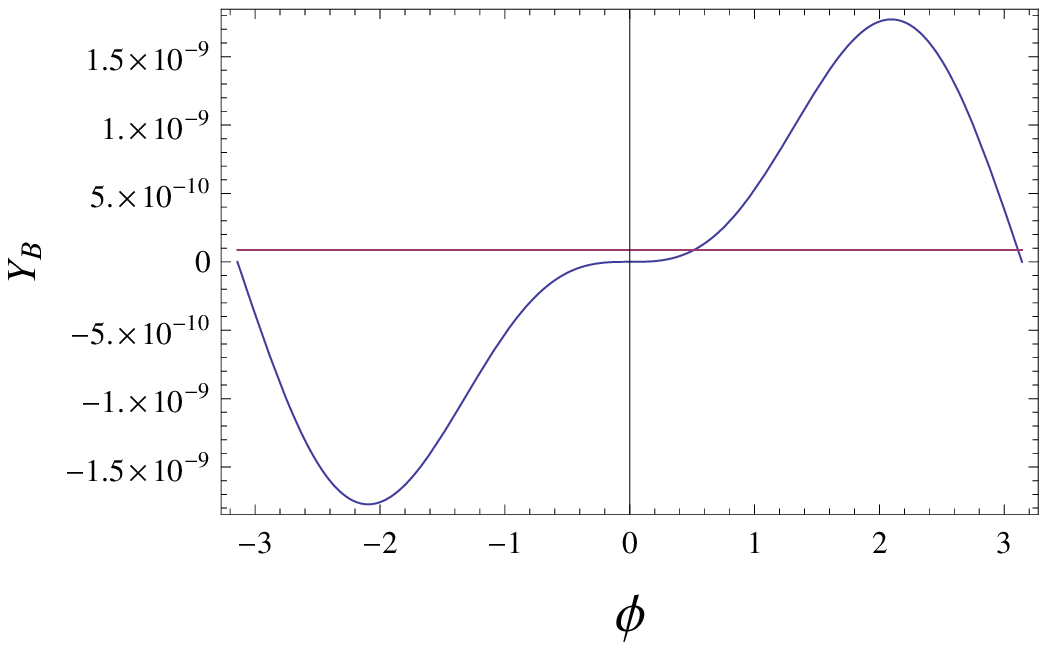}}
\caption{
The resultant baryon asymmetry 
 as a function of the CP-phase $\phi_3$, 
 along with the observed value 
 $Y_B=0.87 \times 10^{-10}$ (horizontal line). 
}
\label{fig:ybIH}
\end{center}
\end{figure}

Fig.~\ref{fig:IHeps} shows the CP-asymmetry parameters,
 $\epsilon_1$ and $\epsilon_2$, as a function of $r$ 
 for $\phi_3=0.1$. 
Two curves are well-overlapped and the peak appears  
 around $r \simeq 10^{-13}$. 
Then, we solve the Boltzmann equations for $r = 10^{-13}$, 
 and show the results in Fig.~\ref{fig:ybIH}.
Although the observed baryon asymmetry in the universe 
 is generated for $\phi_3=0.43 > 0.1$, 
 the neutrino oscillation parameters corresponding to 
 the CP-phase are outside of 2-$\sigma$ range. 
Therefore, in the present scheme, 
 the IH case cannot reproduce the neutrino oscillation 
 data and the observed baryon asymmetry simultaneously.

\section{Conclusions}


In this paper, we have investigated a possibility to explain 
 the baryon asymmetry of the universe 
 as well as the neutrino oscillation data 
 in the TeV scale $B-L$ model.
In the model, the lepton asymmetry is generated in the early universe
 via out-of-equilibrium decays of the right-handed neutrinos 
 with the CP-asymmetry parameter, and  converted 
 into the baryon asymmetry via the sphaleron process. 
When the mass scale of the right-handed neutrinos  
 is low $\lesssim 10^{10}$ GeV, 
 the enhancement of the CP-asymmetry parameter 
 is crucial in order to generate sufficient amount 
 of baryon asymmetry in the universe. 
The enhancement is realized when two right-handed neutrinos 
 are almost degenerated and in this case, 
 the CP-asymmetry parameter can be in principle order unity. 
This scenario is called the resonant leptogenesis. 
However, it is still non-trivial whether the resonant leptogenesis  
 can realize the observed baryon asymmetry 
 in the context of the minimal $B-L$ model, 
 because the $B-L$ interaction mediated by the $Z'$ boson 
 can dramatically reduce the generation of baryon asymmetry. 

We numerically solved the Boltzmann equations 
 for the resonant leptogenesis in the minimal $B-L$ model, 
 and figured out the response between the generated baryon asymmetry 
 and the model-parameters such as the neutrino Dirac Yukawa couplings 
 and the right-handed neutrino masses. 
We  first analyzed the Boltzmann equations with only one-flavor 
 right-handed neutrino and a fixed CP-asymmetry parameter. 
When the neutrino Dirac Yukawa coupling is small, $y_D \lesssim 10^{-10.5}$, 
 the amount of the baryon asymmetry becomes larger 
 as the Dirac Yukawa coupling is raised. 
In this parameter region, the $Z'$ boson and Majorana  Yukawa
 coupling effects dramatically suppress the generation of baryon asymmetry. 
For a large Dirac Yukawa coupling, the baryon number generation 
 by the right-handed neutrino decay dominates over 
 the suppression by the $Z'$ boson effect. 
However, a too large Dirac Yukawa coupling in turn 
 suppresses the generation of the baryon asymmetry 
 by the washing-out effect via the inverse-decay process. 
Next, we have analyzed the Boltzmann equations 
 with two-flavor right-handed neutrinos and 
 shown that two-flavor analysis can be essential in general cases. 
With these analyses, 
 we have shown that in some areas of the parameter space
 a sufficient amount of the baryon asymmetry can be generated 
 though the resonant leptogenesis in the TeV-scale $B-L$ model.

Finally, we have checked whether these parameters are
 consistent with the current neutrino oscillation data. 
We have introduced a simple ansatz for the neutrino  mass matrices, 
 by which the neutrino Dirac Yukawa couplings are 
 determined as a function of a single  CP-phase. 
For both the normal hierarchical and inverted-hierarchical 
 mass spectra of the light neutrinos, 
 we have shown the correlations between the neutrino oscillation 
 parameters and the generated baryon asymmetry via the resonant leptogenesis. 
In our analysis with the ansatz, 
 a fixed CP-phase can reproduce simultaneously 
 the neutrino oscillation data and the observed baryon asymmetry 
 in the normal hierarchical case. 
On the other  hand, we cannot find such a CP-phase 
 in the inverted hierarchical case.

\section*{Acknowledgments}
The work of N.O. is supported in part 
 by the DOE Grants, \# DE-FG02-10ER41714.
Y.O. would like to thank Department of Physics and Astronomy, 
 University of Alabama for hospitality during his visit.

\newpage
\noindent{\Large \bf Appendix}
\appendix

The number density $n_\psi$ of a particle $\psi$ 
 (a right-handed neutrino in our case) 
 with mass $m_\psi$ in the early universe is evaluated 
 by solving the Boltzmann equation of the form \cite{KT}, 
\begin{eqnarray}
 \frac{dY_\psi}{dz}=-\frac{z}{sH(m_\psi)}\sum_{a,i,j,\cdots}
 \left[\frac{Y_\psi Y_a \cdots}{Y_\psi^{eq} Y_a^{eq} \cdots}
 \gamma^{eq}(\psi+a+\cdots\rightarrow i+j+\cdots)\right.\nonumber\\
 \left.-\frac{Y_i Y_j\cdots}{Y_i^{eq} Y_j^{eq}\cdots}
 \gamma^{eq}(i+j+\cdots\rightarrow\psi+a+\cdots)\right], 
\label{boltzeq}
\end{eqnarray}
where $Y_\psi =n_\psi/s$ is the ratio of $n_\psi$ 
 and the entropy density $s$, $z=\frac{m_\psi}{T}$, 
 and $H(m_\psi)$ is the Hubble parameter at a temperature $T=m_\psi$. 
The right hand side of Eq.~(\ref{boltzeq}) describes 
 the interactions that change number of $\psi$,  
 and $\gamma^{eq}$ is the space-time density of scatterings 
 in thermal equilibrium. 
For a dilute gas we take into account 
 decays, two-particle scatterings and the corresponding back reactions. 
One finds, for a decay the particle $\psi$, 
\begin{equation}
 \gamma_D=\gamma^{eq}(\psi\rightarrow i+j+\cdots)
=n_\psi^{eq}\frac{K_1(z)}{K_2(z)}\tilde{\Gamma}_{rs}, 
\end{equation}
where $K_1$ and $K_2$ are the modified Bessel functions,  
 and $ \tilde{\Gamma}_{rs}$ is the decay width. 
For two body scattering one has 
\begin{equation}
 \gamma\left(\psi+a\leftrightarrow i+j+\cdots\right)
 =\frac{T}{64\pi^4}\int^\infty_{(m_\psi+m_a)^2} ds 
 \hat{\sigma}(s)\sqrt{s}K_1\left(\frac{\sqrt{s}}{T}\right), 
\end{equation}
where $s$ is the squared center-of-mass energy and the reduced cross 
 section $\hat{\sigma}$ for the process 
 $\psi+a\leftrightarrow i+j+\cdots$ is related 
 to the usual total cross section $\sigma(s)$ by 
\begin{equation}
 \hat{\sigma}(s)=\frac{8}{s}\left[(p_\psi\cdot p_a)^2
-m_\psi^2 m_a^2\right]\sigma(s). 
\end{equation}

In the following we list the explicit forms 
 of the reduced cross sections used in our analysis \cite{boltz}. 
The reduced cross section corresponding to $\gamma_N$ is given by 
\begin{eqnarray}
 \hat{\sigma}_N(s)=\frac{\alpha^2}{\sin^4\theta}\frac{2\pi}{M_W^4}\frac{1}{x}
\left[a_1\left(m_D m_D^\dagger\right)_{11}^2
\left(x+\frac{2x}{D_1(x)}+\frac{x^2}{2D_1^2(x)}
\right.\right.\nonumber\\
\left.\left.-\left(1+2\frac{x+1}{D_1(x)}\right)\ln\left(x+1\right)
\right)\right], 
\end{eqnarray}
where $x=\frac{s}{m_N^2}$, and 
\begin{equation}
 D_1(x)=x-1+\frac{c}{x-1},\ {\rm with}\ 
c=\left(\frac{\tilde{\Gamma}_{rs}}{m_N}\right)^2,  
\end{equation}
while the one corresponding to $\gamma_{N,t}$ is 
\begin{eqnarray}
 \hat{\sigma}_{N, t}(s)=&\frac{2\pi\alpha^2}{M_W^4\sin^4\theta}\left[
\left(m_D m_D^\dagger\right)^2_{11}\left(\frac{x}{2(x+1)}
+\frac{1}{x+2}\ln\left(x+1\right)\right)\right]. 
\end{eqnarray}
The reduced cross section for the $t$-channel (and $u$-channel) 
 process $N+N\rightarrow\Phi+\Phi$ mediated by the right-handed 
 neutrino is given by 
\begin{equation}
 \hat{\sigma}_{N, t, \Phi}(s)=\frac{y_N^4}{8\pi}\frac{x-4}{x}
  \left(-2+\frac{x}{2}+\frac{x^2-8x+16}{x\sqrt{x(x-4)}}
  \log\frac{x-\sqrt{x(x-4)}}{x+\sqrt{x(x-4)}}\right). 
\end{equation}
The reduced cross section for the $s$-channel process 
 $N+l\rightarrow\bar{t}+q$ mediated by the Higgs doublet 
 is given by 
\begin{equation}
 \hat{\sigma}_{h, s}(s)=\frac{3\pi\alpha^2m_t^2}{M_W^4\sin^4\theta}
\left(m_D m_D^\dagger\right)_{11}\left(\frac{x-1}{x}\right)^2,  
\end{equation}
 while for the $t$-channel process 
\begin{equation}
 \hat{\sigma}_{h, t}(s)=\frac{3\pi\alpha^2m_t^2}{M_W^4\sin^4\theta}
\left(m_D m_D^\dagger\right)_{11}\left[\frac{x-1}{x}
+\frac{1}{x}\ln\left(\frac{x-1+y^\prime}{y^\prime}\right)\right]  
\end{equation}
 with $y^\prime=\frac{m_h^2}{M_1^2}$. 
The total reduced cross section for the process 
 $f+\bar{f}, \Phi+ \Phi\rightarrow N+N$ mediated 
 by the $Z'$ boson ( $f$ denotes the SM fermions) 
 is given by 
\begin{equation}
 \hat{\sigma}_{Z^\prime}(s)=\frac{104\pi}{3}\alpha_{B-L}^2
\frac{\sqrt{x}}{(x-y)^2+yc}\left(x-4\right)^{\frac{3}{2}},  
\end{equation}
 where $y=\frac{m_{Z'}^2}{m_N^2}$, and 
 $c=\left(\tilde{\Gamma}_{Z^\prime}/{M_1}\right)^2$ 
 with the decay width of $Z'$ boson 
\begin{equation}
\tilde{\Gamma}_{Z^\prime}=\frac{\alpha_{B-L} m_{Z^\prime}}{6}
\left[ 3\left(1-\frac{4}{y}\right)^\frac{3}{2}\theta(y-4)
+13\right]. 
\end{equation}


\end{document}